\documentclass[12pt]{article}
\usepackage{colordvi,amssymb,color}
\usepackage[dvips]{graphicx}
\begin{document}
\begin{center}
{\Large \bf Shadowing effects for continuum and discrete deposition
models\\}
\vspace*{.5cm}
{\normalsize T H Vo Thi$^1$, P Brault$^2$,  J L Rouet$^{1,3}$ and  S Cordier$^1$}
\vspace*{.5cm}

$^1$ Laboratoire de Math\'ematiques et Applications, Physique
Math\'ematique d'Orl\'eans, UMR6628
CNRS-Universit\'e d'Orl\'eans, BP 6759, 45067 Orl\'eans Cedex 2, France \\
\vspace*{.15cm}

$^2$ Groupe de Recherches sur l'Energ\'etique des Milieux Ionis\'es,
UMR6606 CNRS-Universit\'e d'Orl\'eans, BP 6744, 45067 Orl\'eans Cedex 2, France\\
\vspace*{.15cm}

$^3$ Institut des Sciences de la Terre d'Orl\'eans
UMR6113 CNRS/Universit\'e d'Orl\'eans, 1A rue de la F\'erollerie, 45071 ORLEANS
CEDEX 2
\vspace*{.5cm}

Email~: ThuHuong.VoThi@univ-orleans.fr, Pascal.Brault@univ-orleans.fr,
Jean-Louis.Rouet@univ-orleans.fr,
Stephane.Cordier@univ-orleans.fr
\end{center}

\begin{abstract}
We study the dynamical evolution of the deposition interface using both discrete and
continuous
 models for which shadowing effects are important. We explain why continuous and
discrete models 
implying both only shadowing deposition do not give the same result and propose a
continuous model 
which allow to recover the  result of the discrete one exhibiting a strong columnar
morphology.
\end{abstract}
\section{Introduction}
Shadowing during thin film deposition is an important effect resulting in a
particular surface roughness evolution \cite{bales,karu,roland}. In that case thin
films exhibit a columnar structure. Shadowing results in columns due to masking of
incoming particles by elevated part of the surface. In that case higher values of
the surface are expected to grow faster than lower ones \cite{bales,karu}.\\

Two approaches have been proposed to model the deposition process : models based on
probabilistic methods \cite{tombo,meng,ficht} handled with Monte-Carlo (MC) methods
and continuous models based on partial derivative equations (PDE)
\cite{sarma,rost,amar}. Deposition is usually characterized by the root mean square
of the fluctuations $W(L,t)$ of the interface height $h(x,t)$ at position $x$ and
time $t$. $L$ is the length of the simulated system. Dynamical scaling hypothesis
implies the relation $W(L,t)=L^{\alpha}f(t/L^{\alpha/\beta})$. Where $f$ is a
scaling function such that $f(y)\sim y^{\beta}$,  for small $y$, in order to give a $W$ independent
of $L$ at small time and  $f(\infty) \sim const$, which describes a saturation of $W$
because of size effects\cite{tong,bara,witten,lagues}. The values of the exponents
$\alpha$ and $\beta$ depend on the model. Discrete or continuous models which
exhibit the same scaling exponents belong to the same universality class and are
expected to describe the same deposition process.  For example, numerical 
simulations show that the Edward-Wilkinson model and the Random Deposit Relaxation
belongs to the same universality class \cite{bara,lagues,edward}. One  should
mention that in the case of the restricted solid on solid (RSOS) model a
correspondence has been established with the KPZ model \cite{park,kardar}. \\

In this work, we examine the properties of the solution of both continuous and
probabilistic models which describe columnar growth especially occurring in plasma
sputter deposition process. The columnar shape is mainly due to shadowing effects
and has been described with MC and PDE models. One goal of this paper is to
highlight the correspondence between the two descriptions 
in the case of a columnar deposition type. In this respect, the root mean square
fluctuation $W(L,t)$ is not enough to characterize the deposit and we will use the
height distribution function.\\

By contrast with MC models, continuous models are not known to  produce columnar
deposit shape. To that respect the shadowing model proposed by Yao {\it et al}
\cite{yao1,yao} and Drotar {\it et al} \cite{drotar} are first steps toward this
direction. The continuous model only gives columns at early time while at long time a
few sharp peaks remain due to shadowing. Moreover their interfaces, while having strong
differences between the minimal and maximal height, do not present 
strong slopes as MC solutions do.\\

 In this work, we will focus, in section two, on shadowing discrete and continuous
model and their respective results. We especially highlight why there is a
difference between Monte-Carlo and continuous models including shadowing and
deposition. Then, section three, we present a new continuous non-local nonlinear
model  which allow to recover the results based on the Monte-Carlo technique
described in section two, preserving columnar growth even for large times. Section 4 provide discussion
and conclusion.

\section{Models with shadowing}

Because it has been recognized that the shadowing effects drive to the formation
 of
columnar deposit, we study both continuous and discrete models only considering this
sole aspect \cite{roland,yao}.\\

\begin{figure}[!h]
\centerline{\includegraphics*[width=0.4\linewidth]{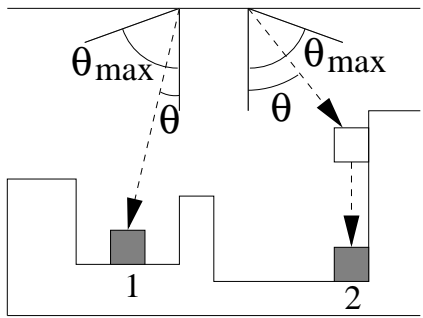}}
\caption{\label{MC1-1} \it Discrete model. SOS shadowed ballistic deposition
process. Particle 2 hits a column side, then falls down.}
\end{figure}

A discrete model has been proposed by Roland and Guo \cite{roland}. The substrate
and the source are parallel and are at large distance from each other. The main idea
of this model is to release particles from the source with an
angle $\theta$, measured with respect to the surface normal and taken at random
between $\pm \theta_{max}$. Then particles have ballistic trajectories unless they
strike the interface (see particle 1 in figure \ref{MC1-1}). If the particle hits
the side of an existing column, it falls down (see particle 2 in figure
\ref{MC1-1}). This condition is in agreement with an SOS model for which overhanging
is forbidden. Obviously, it is also possible to introduce surface relaxation
effects driven by the surface temperature using Arrhenius law \cite{meng}. Because
we focus on the shadowing effects, we will not take into account relaxation effects.
Except the relaxation, this model is the one proposed by Roland and Guo
\cite{roland}.

\begin{figure}[!h]
\centerline{\includegraphics*[angle=-90,width=0.6\linewidth]{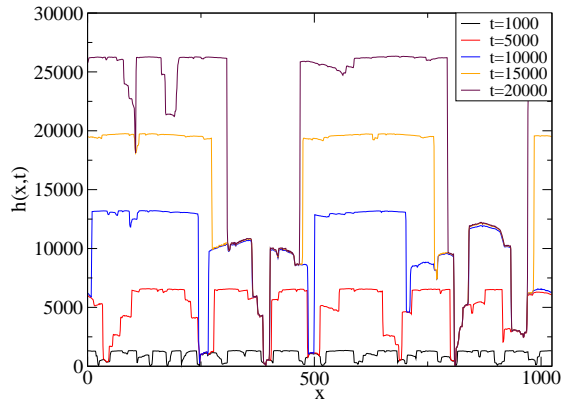}}
\caption{\label{MC1-2}\it Discrete model.  Interface $\,h(x,t)\,$ at time
$t=1\,000$, 5\,000, 10\,000, 15\,000 and 20\,000 for $\theta_{max}=60^{\circ}$ and
$L=1024$.}
\end{figure}

\begin{figure}[!h]
\centerline{\includegraphics*[angle=-90,width=0.6\linewidth]{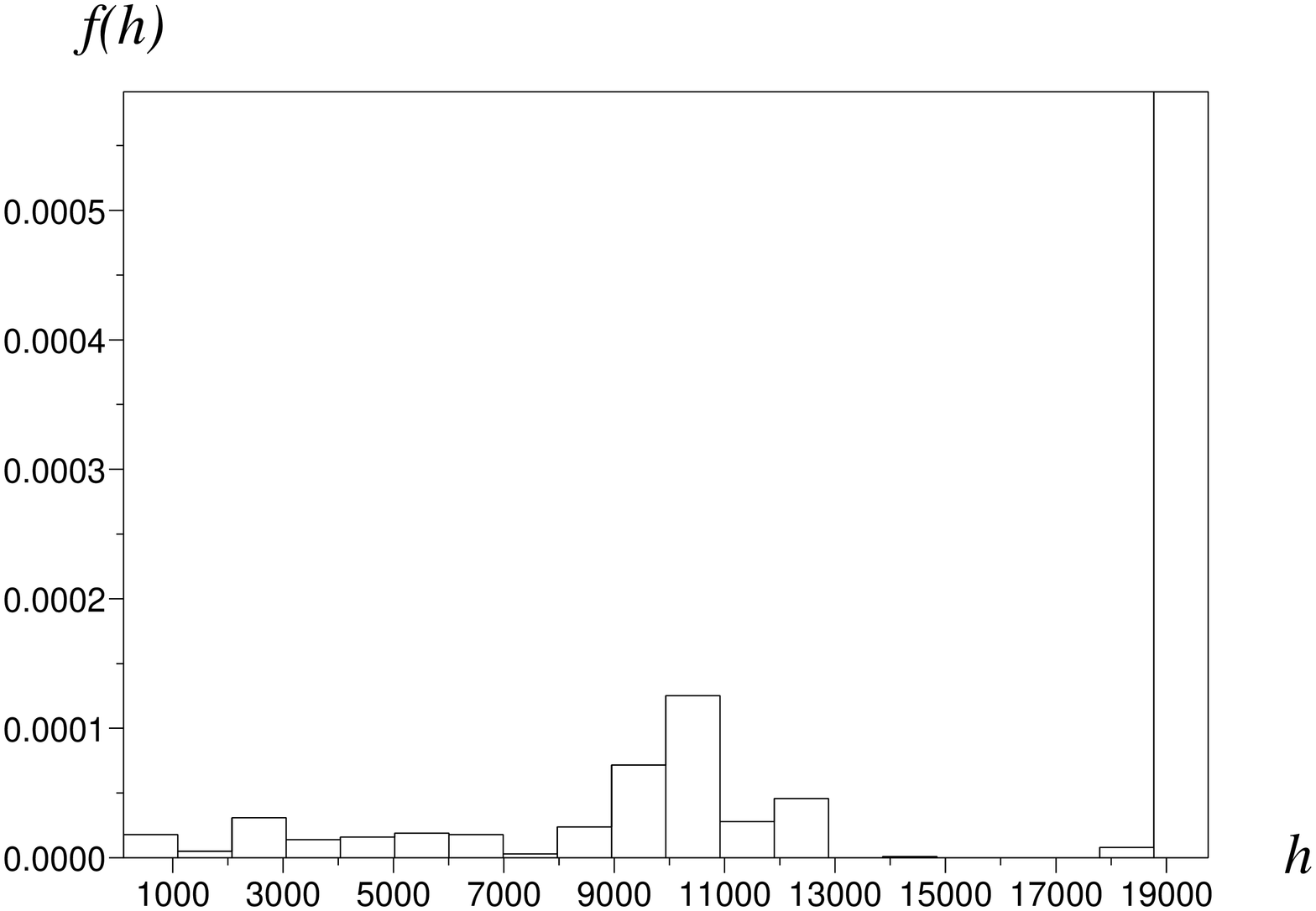}}
\caption{\label{MC1-3} \it Distribution function of the interface height $h(x,t)$ plotted figure
\ref{MC1-2} at time $t=15\,000$.}
\end{figure}

\begin{figure}[!h]
\centerline{\includegraphics*[angle=-90,width=0.6\linewidth]{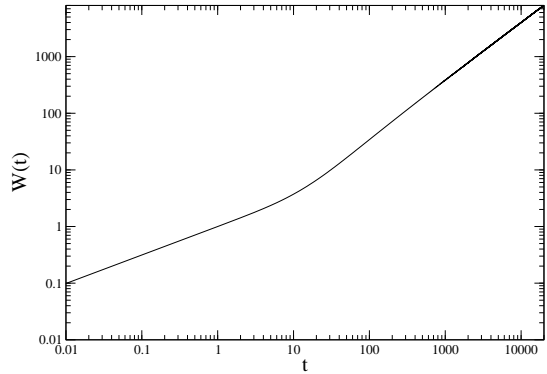}}
\caption{\label{MC1-4} \it Time evolution of the roughness $W(t)$ for the simulation
the results of which are given figure \ref{MC1-2}.}
\end{figure}

A typical result of this model is given  figure \ref{MC1-2}. It is obtained for a
periodic system of length $L=1024$ and $\theta_{max}=60^{\circ}$. Figure \ref{MC1-2}
gives the value $h(x,t)$ of the interface at different times. We observe that the
system forms height flat structures with deep grooves in between. The column width
increases with time while the number of column decreases.  This columnar shape grows
and persists at least until the end of the simulation. Figure \ref{MC1-3} gives the
distribution $f(h)$ of the height of the interface at time $t=20\,000$.
It exhibits a single large peak for $h\sim19\,000$ which corresponds to
the top of the plateau, showing its flatness. The smallest values of $f(h)$
correspond to the bottom of the grooves which indeed remain at low values. This
curve shape characterizes the columnar regime. Nevertheless it do not gives any
information on the number of columns. It could be evaluated as in \cite{yao1}.
Because it is not the main interest of the paper, we turn now to the roughness $W(t)$, given
figure \ref{MC1-4}, which is a more usual function in the deposition context. It
exhibits two regimes. The first one until $t\sim 10$ for which $W\sim t^{.5}$
corresponds to a fluctuating interface. The second one for which $W\sim t$ corresponds
to the columnar regime.\\

It has been recognized that, if a relaxation term is introduced, the columnar aspect
is less strong and 
completely disappear if it is too high \cite{yao}.\\

\begin{figure}[!h]
\centerline{\includegraphics*[width=\linewidth]{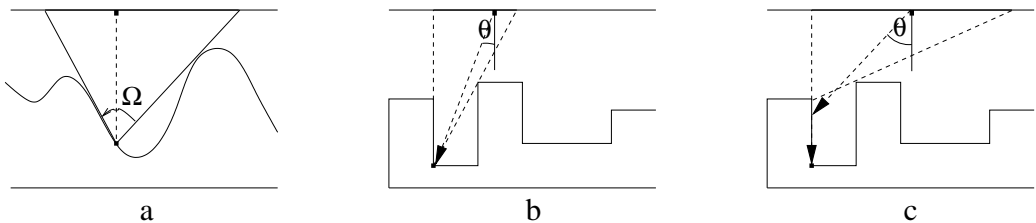}}
\caption{\label{MC2-1} \it Shadowing process : comparison between continuous (a) and
discrete (b,c) models. Discrete model : the point at the bottom of a column receives particles if it
is seen by the target (b) or if the side of the column is seen by the target (c).}
\end{figure}

Now let us turn to a continuous model. The following model, which include shadowing
effects, has been proposed by Karunasiri et al \cite{karu}.

\begin{equation}
\frac{\partial h}{\partial t}=\nu \nabla^2 h+\frac{\lambda}{2}(\nabla h)^2 +
R\Omega(x,\{h\}) + \eta
\label{eq3.0}
\end{equation}

which is a KPZ equation for which the deterministic deposition term $R$ is
multiplied by the solid 
angle $\Omega(x,\{h\})$ which modelizes the
shadowing effect. $\Omega(x,\{h\})$ is the solid angle subtended by the target surface seen
from a point $x$ on the interface height $h(x,t)$ (see figure \ref{MC2-1}(a)). It is
evaluated as in reference \cite{karu}. $\nu$ is the diffusion coefficient which is
constant and $\eta$ the usual noise the mean of which, at time $t$, is equal to zero
and the correlation given by $<\eta(x,t)\eta(x',t')>=2D\delta(x,x')\delta(t,t')$.
Because we are only interested in the shadowing term effect, we reduce the
analysis to this sole aspect, as we do for the Monte-Carlo simulations and
consequently study the following equation~:

\begin{equation}
\frac{\partial h}{\partial t}=R\Omega(x,\{h\}) + \eta
\label{eq3}
\end{equation}

Equation (\ref{eq3}) is numerically integrated using a finite difference method with
an explicit scheme. The time step $\Delta t=0.05$, the spatial step $\Delta x=1$, $R=1$ and we use
periodic boundary conditions. At each integration time step the exposure angle
$\Omega$ is compute for each point, then normalized with the mean value obtained at
that time. This insure that the deposition rate is indeed constant and that each
unit of time a single layer is growing on average.\\

\begin{figure}[!h]
\includegraphics*[angle=-90,width=0.31\linewidth]{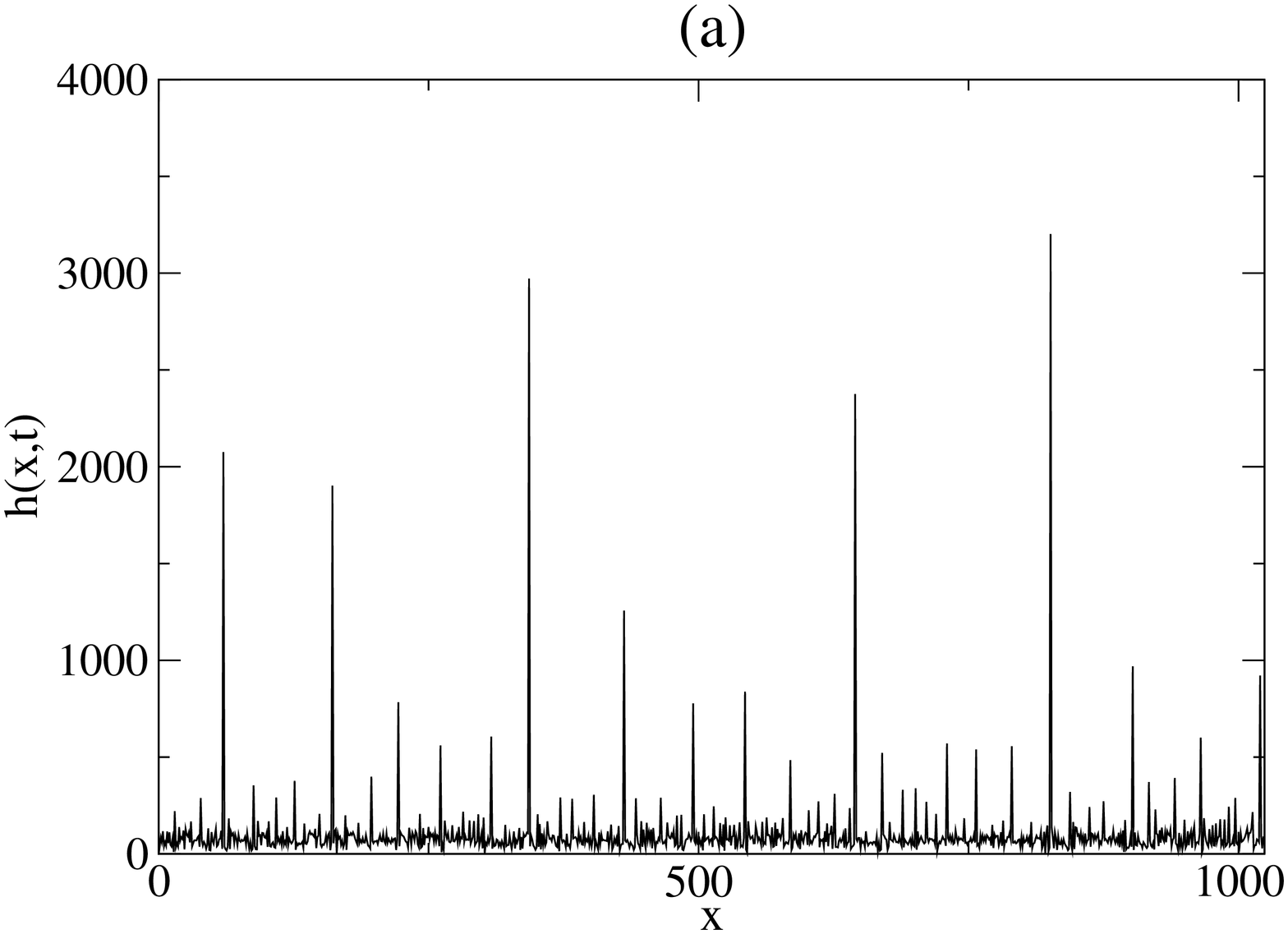}
\includegraphics*[angle=-90,width=0.31\linewidth]{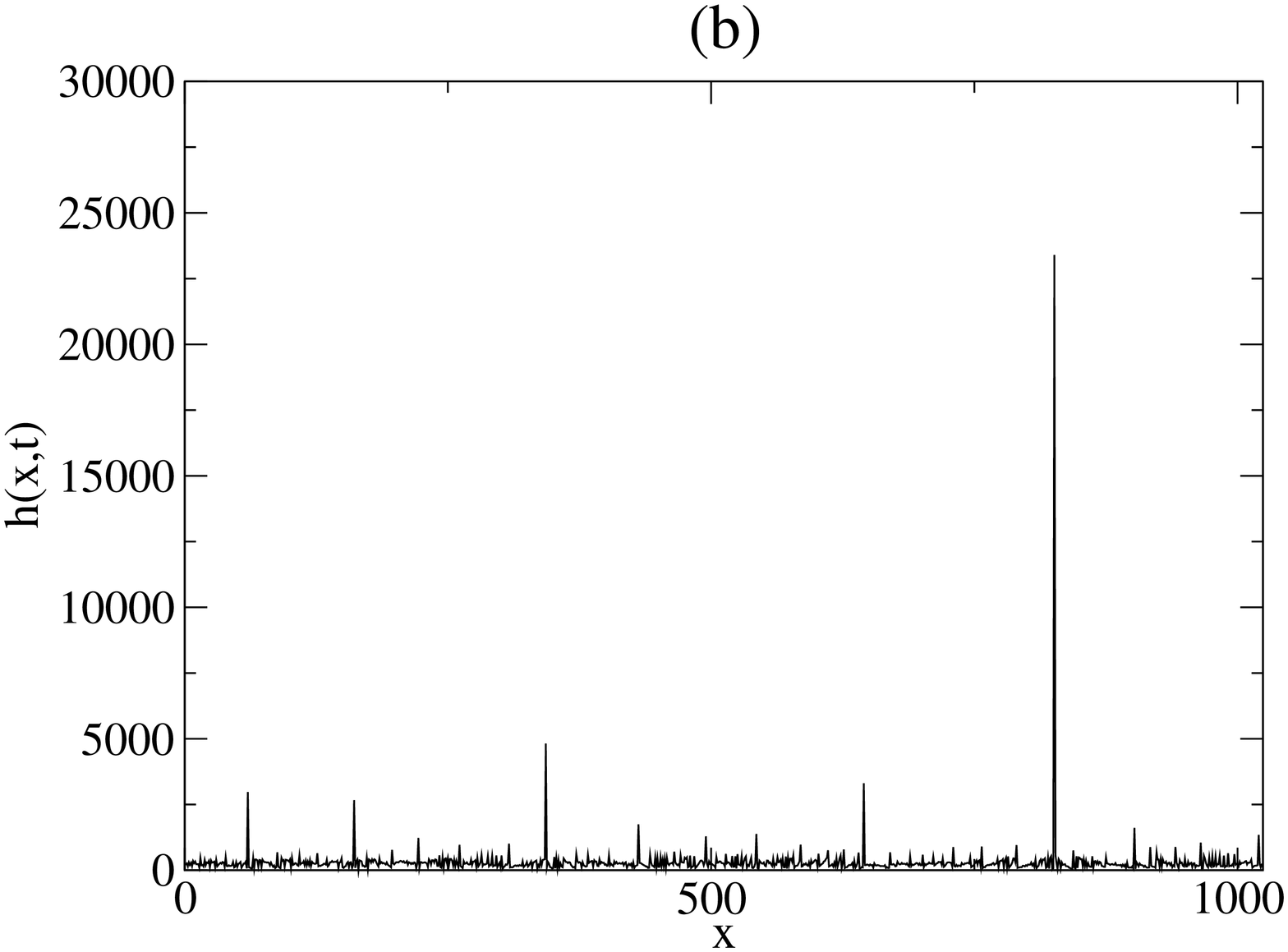}
\includegraphics*[angle=-90,width=0.31\linewidth]{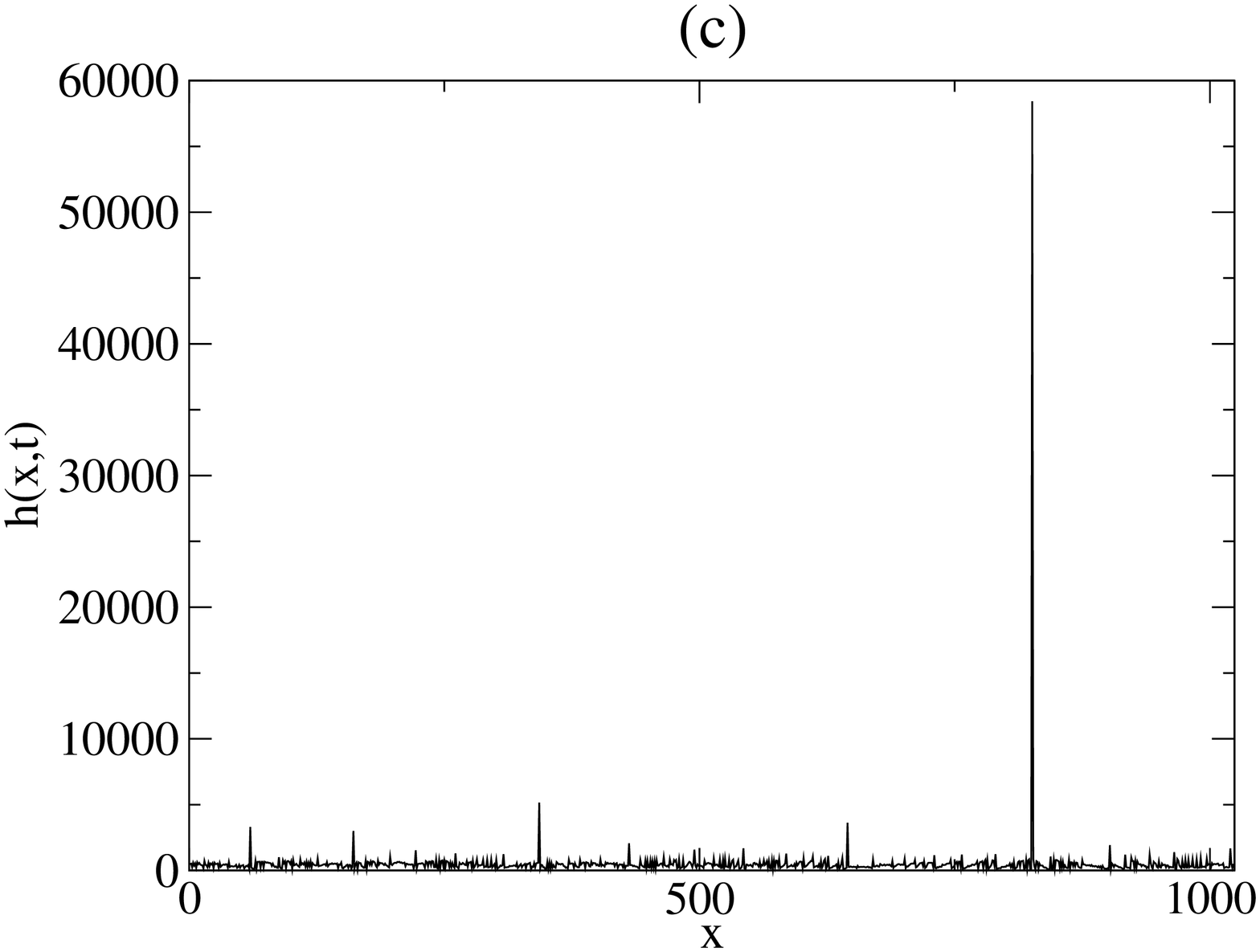}
\caption{\label{EDP1-1} \it Interface $\,h(x,t)\,$ for a continuous model with
shadowing deposition at time $t=100$ (a), $t=300$ (b), $t=500$ (c).}
\end{figure}

Figures \ref{EDP1-1} gives the surface morphologies obtained at time $t=100$, $200$
and $300$ for a surface length $L=1024$. Indeed the shape of $h(x,t)$ is very
different from the one obtained with the discrete model previously discussed. We do
not have plateau but instead very high peaks. As the simulation goes on, a single
peak emerges. Nevertheless this last effect is due to the finite size of the
system.\\

In order to improve continuous models, this difference has to be explained. For
these models, the height of a site increases proportional to its exposure angle.
This angle reaches $\pi$, and is maximum for the highest sites. By contrast,
small height sites received less particles. More the difference is, less
important is the increase rate for these sites. For example at time $t=500$, all
points (except the single highest peak)  have small exposure angle and then do not
grow up very much while the peak quickly increases. The situation is the same for
the discrete model except for sites which are at the bottom of a column. In fact for
these sites, particles come directly from the source if they are in the correct
exposure angle  (figure \ref{MC2-1}(b)) but also fall down from the side of the column (see figure
\ref{MC2-1}(c)).\\

\begin{figure}[!h]
\includegraphics*[angle=-90,width=0.31\linewidth]{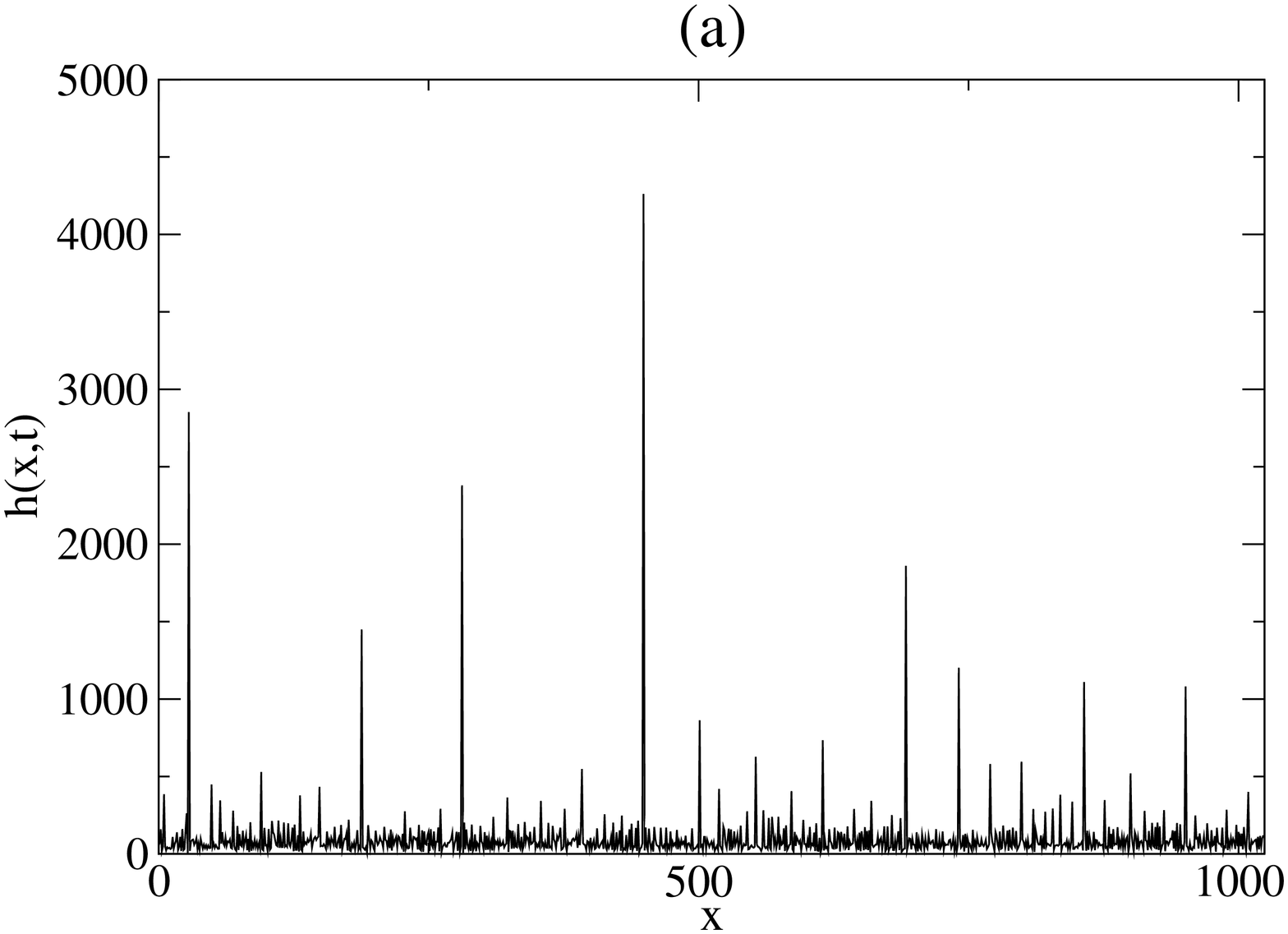}
\includegraphics*[angle=-90,width=0.31\linewidth]{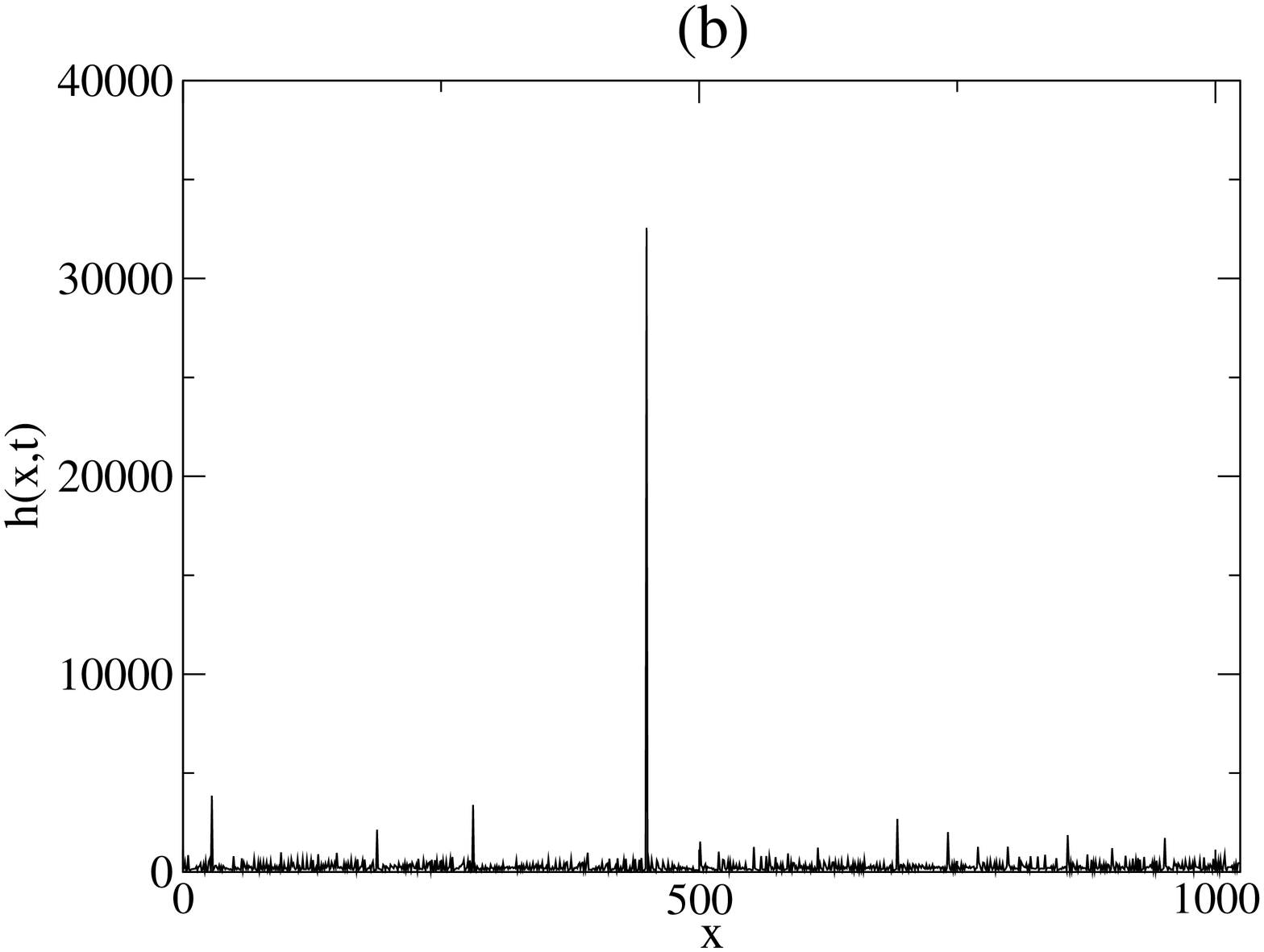}
\includegraphics*[angle=-90,width=0.31\linewidth]{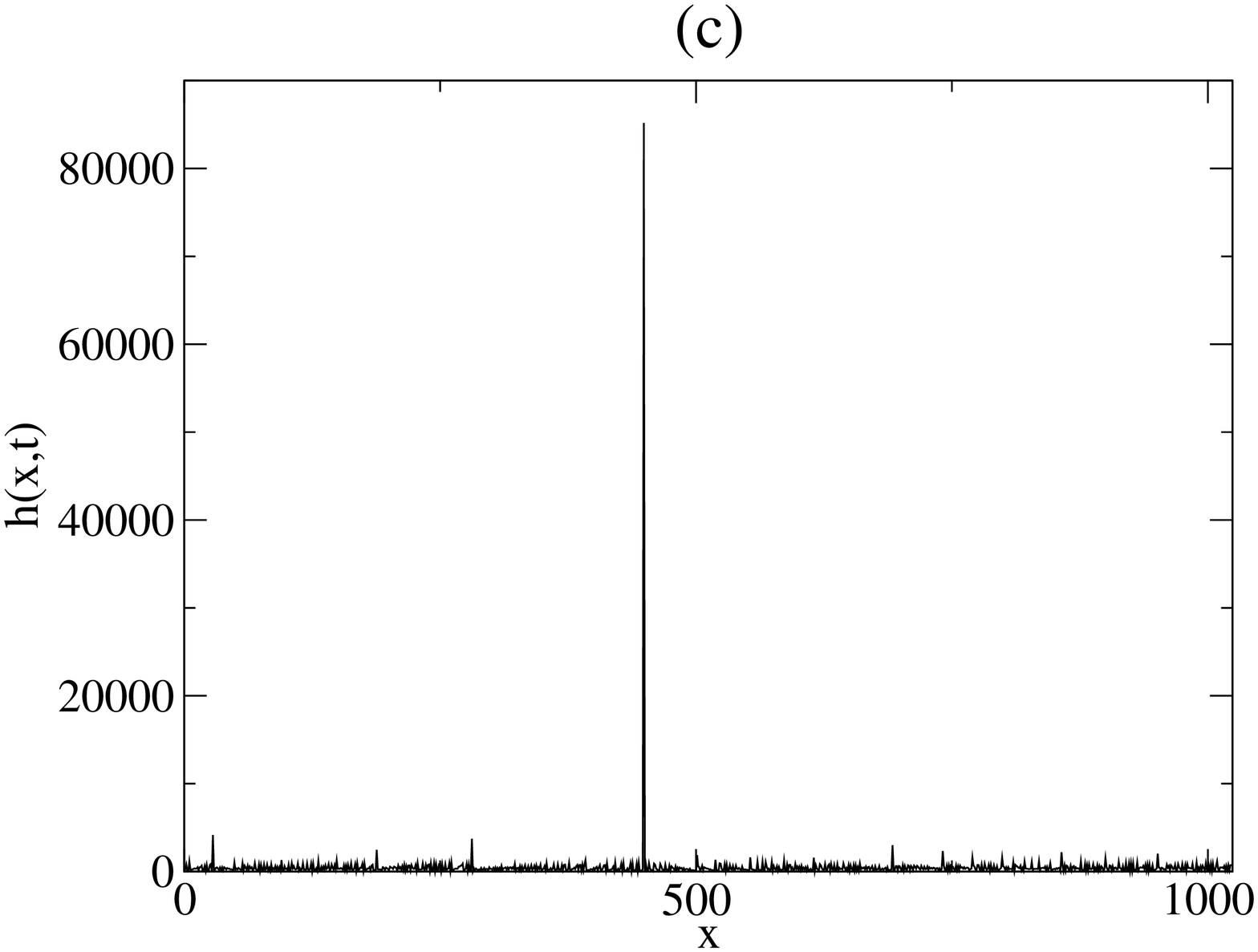}
\caption{\label{MC2-2} {\it Discrete model. Interface height for a periodic system
of length  $L=1024$, at time $t=100$ (a), $t=300$ (b), $t=500$ (c). Particles which
are hitting the side 
of a column are removed.}}
\end{figure}

It is possible to check this hypothesis by performing a peculiar discrete model. Now
if a particle hit the edge of a column it is removed and consequently cannot fall
down. Figures \ref{MC2-2} show the height $h(x,t)$ of the interface at time $t=100$,
300 and 500. Indeed these figures are very similar to those obtained with the
continuum model with shadowing (figures \ref{EDP1-1}). Except to get a discrete
model that mimic the continuous one, there is no physical reason for that process.
Nevertheless it shows that the shadowing introduced in the continuous model is not
enough to produce columnar shape deposit, and in addition it also shows the
importance of the particle flux which fall down from the edges of the columns. This
suggests the introduction of a stronger relaxation at the top of the
column than at their bottom in the continuous model.\\

\section{Continuous model driving to columnar shape deposit}

We propose the following stochastic differential equation where
the main ingredients are a non-linear shadowing effects and a anisotropic
diffusion~:

\begin{equation}
\frac{\partial h}{\partial t}=g(\Omega(x,\{h\}))\,(R\,\sqrt{1+|\nabla(h)|^2}+\nu
\nabla^2 h+ \eta)
\label{huong1}
\end{equation}

In this equation $g(\Omega)$ is a given function of the solid angle $\Omega$. The
square root term $\sqrt{1+|\nabla(h)|^2}$ describes the fact that the local deposit
grows normally to 
the interface. Here we do not make a first order approximation, as in \cite{kardar},
because, as we are looking for columnar shape deposit, the spatial variation of the
height $h$ could be large. The fact that the solid angle, which modelizes the
shadowing,  is in factor the the right hand term. It
will obviously increase the deposit rate for surfaces which are not shadowed (mainly
for large value of $h$) and make it smaller for shadowed one (for small value of
$h$). The diffusion is also affected by the shadowing. It has been recognized in
section 2 that particles which fall down a column side increases the width of this
column. This strong migration of particles
 from the side to the bottom of a column
suggests this anisotropic diffusion. Moreover, in order to increase the shadowing
effect, we take $g(\Omega)=\Omega^2$.\\

Equation (\ref{huong1}) is integrated using the following explicit scheme:

\begin{eqnarray}
h_i^{n+1} & = & h_i^n
                +\left(\Omega_i^n\right)^2\,\left[\Delta
t\,R\,\sqrt{1+\left(\frac{h_{i+1}^n-h_{i-1}^n}{2\,\Delta
x}\right)^2}\right.\\ \nonumber
              &   &  
                + \frac{\nu\,\Delta t}{\Delta x^2}\,(h_{i+1}^n-2\,h_i^n+h_{i-1}^n)
                +\left. \sqrt{\frac{2\,D\,\Delta t}{\Delta x}}\,\,\varepsilon\right] 
\label{huong2}
\end{eqnarray}
with the notation $h_i^n=h(i\,\Delta x, n\,\Delta t)$, $\varepsilon$ is a random
number picked with the uniform distribution between $[-1,1[$.\\

\begin{figure}[!h]
\includegraphics*[angle=-90,width=0.6\linewidth]{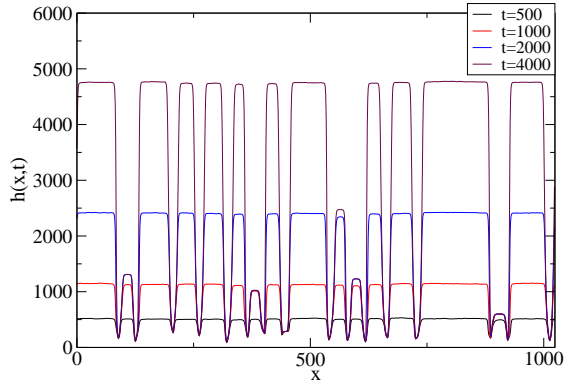}
\caption{\label{shadow1} Continuous model. Snapshot of the interface given by the
non-linear shadowing anisotropic
 diffusion model given by equation (\ref{huong1}) at time $t=500$, 1\,000, 2\,000
and 4\,000.}
\end{figure}

\begin{figure}[!h]
\includegraphics*[angle=-90,width=0.6\linewidth]{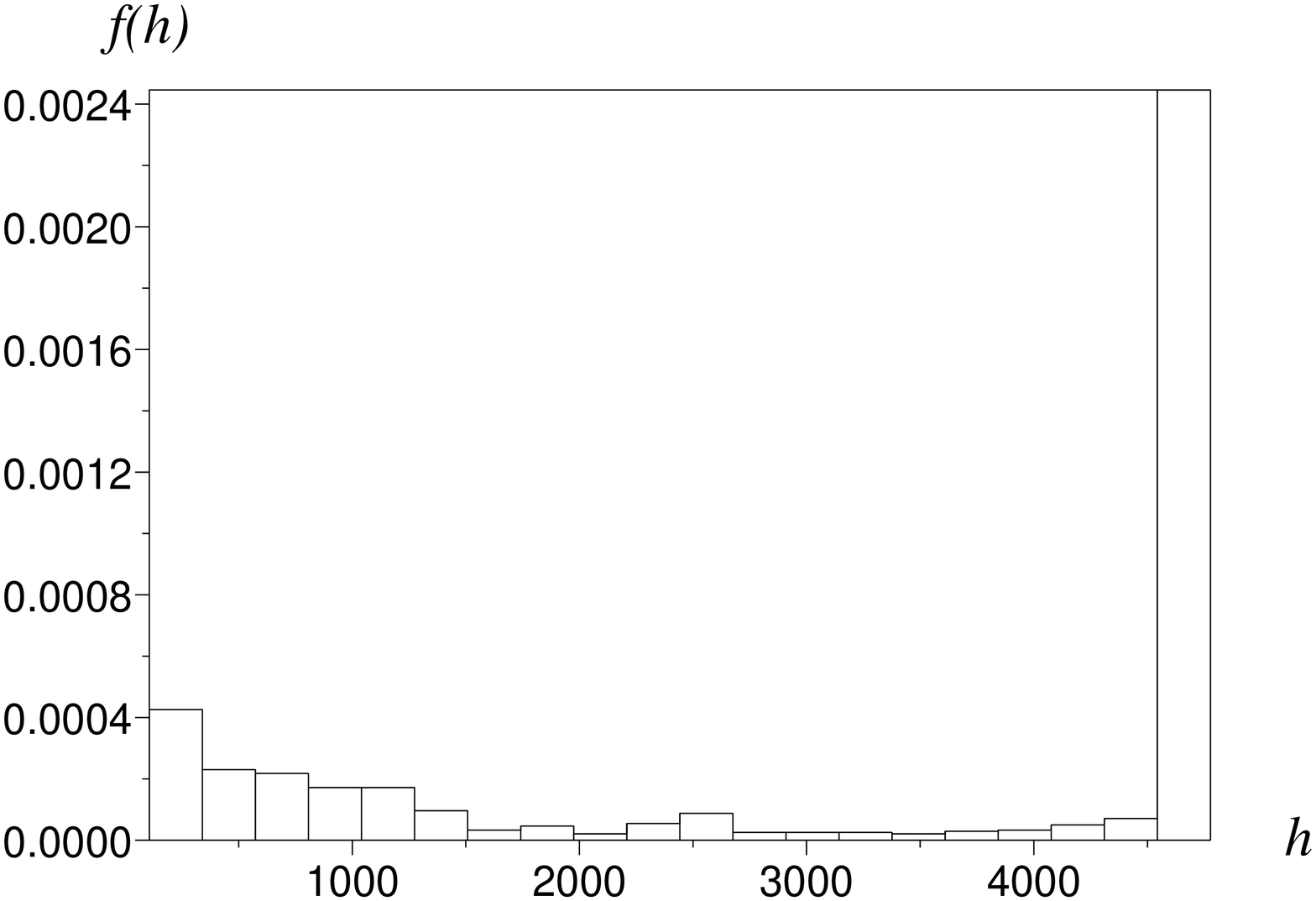}
\caption{\label{shadow2} \it height distribution function of the interface
corresponding to figure \ref{shadow1} at $t=4\,000$.}
\end{figure}

\begin{figure}[!h]
\includegraphics*[angle=-90,width=0.6\linewidth]{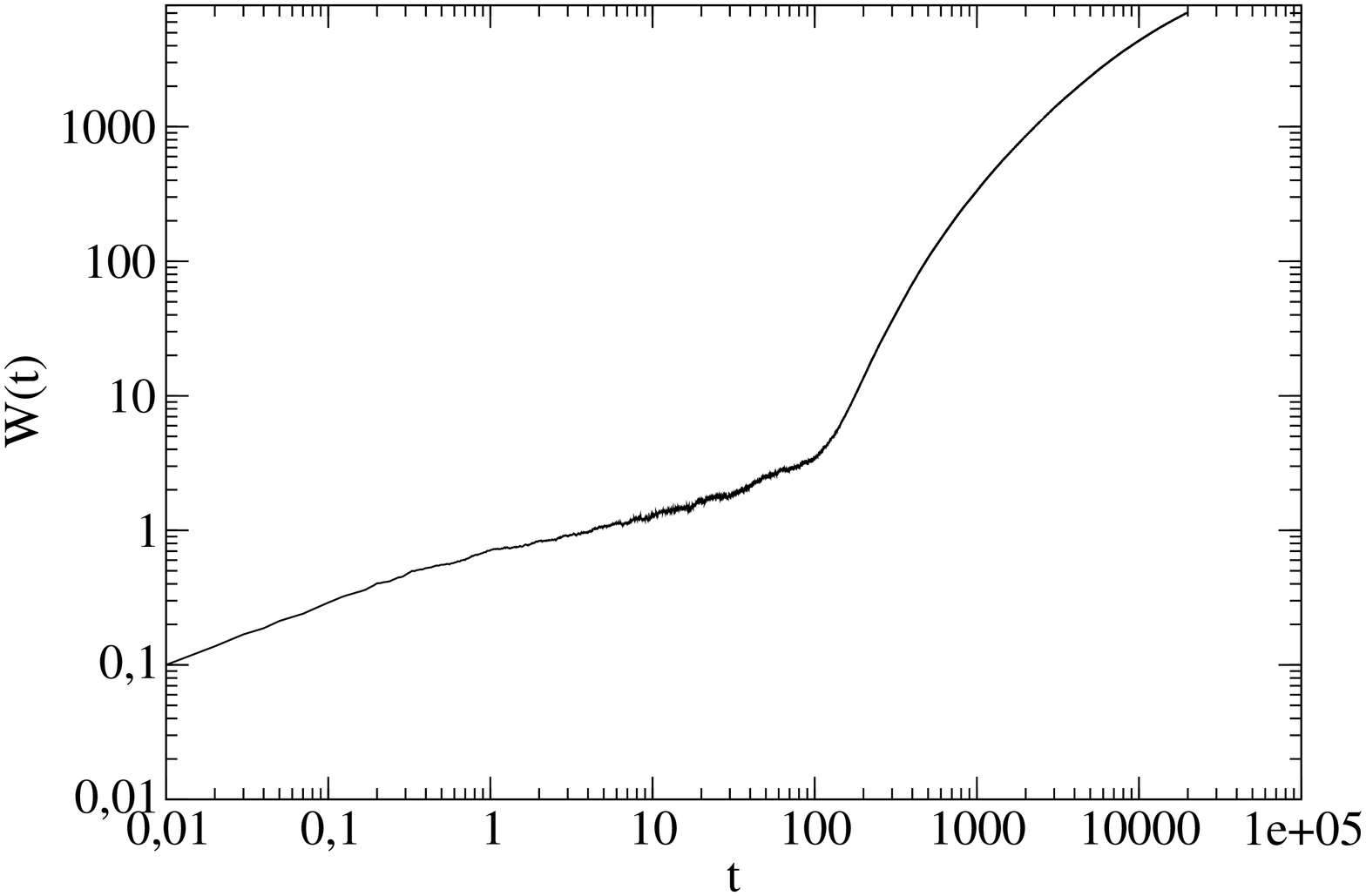}
\caption{\label{shadow3} \it Continuous model. Time evolution of the roughness $W(t)$.}
\end{figure}

A plane wave analysis with a perturbation $h=h_1\exp(i(kx-\omega t))$, on the
linearized version of equation (\ref{huong1}), gives the following dispersion
relation

\begin{equation}
- i\omega=2\frac{\alpha\,R}{\bar{\Omega}}\, k-\nu\,k^2,
\label{huong3}
\end{equation}

where $\bar{\Omega}$ is the mean value of the solid angle and $\alpha \sim .7$. With
$\omega=\omega_R+i\omega_I$, equation (\ref{huong3}) shows that the modes
$k<k_*=2\alpha R/(\bar{\Omega}\nu)$ are unstable. Then, the noise trigger the
instability and drives the system into a strong non-linear regime. Then, as shown
figure \ref{shadow1}, which gives the evolution of the interface profile at
different times for $\Delta t=0.01$, $\Delta x=1$, $D=1$, $\nu=1$ and $R=1$, as time
increases,  a columnar deposit shape is observed. The competition between the
shadowing deposition, which favors the emergence of a single structure (see figure
\ref{EDP1-1}), and the anisotropic diffusion, which propagate particles near the
edges,  keeps at least for the simulation time ($t=4\,000$) the columnar regime.
Indeed figure \ref{shadow1} shows the formation of higher and higher columns.
Moreover, most of the columns formed at the beginning of the simulation are still
present at the end. The height distribution function (figure \ref{shadow2}),
computed at $t=4\,000$, is very similar to the one obtained with the discrete model
and shows a strong peak for $h\sim 5\,000$ which correspond to the top of the columns.
The time evolution of the roughness $W$ of the interface is given figure
\ref{shadow3}. It shows the existence of different regimes. The
first one, for $t<1$ is driven by the fluctuations and $W$ scales as $t^{1/2}$. For
the second one ($1<t<100$), diffusion induced a relative reduction of the roughness
which scales as $t^{.4}$. Then, because of the shadowing instability described above, sharp canyons appear and the roughness quickly
increases. Finally, after $t\sim 1000$, the columnar regime appears and gives
$W\sim t$ as in the discrete model.\\

\section{Discussion and Conclusion}

In the past, both discrete and continuous models have been proposed to describe
columnar shape deposit as those observed in plasma sputter deposition. Discrete
models  implying shadowing process, as the one established by Roland and Guo
\cite{roland}, indeed show the formation of larger and larger plateau as time
increases. They also proposed a continuous model for which, each point of the
interface received a flux of particles proportional to the local exposure angle.
Then the interface obtained presents peaks but no columns as the discrete model do.
We show that these two shadowing processes are not equivalent.  For the discrete
one, points which are at the bottom of columns received all the particles which are
hitting the edges. There is no correspondence of such a process in the current continuous
model. Nevertheless if we remove particles which are hitting the edges in discrete
simulations, then the two types of models, for which only the shadowing deposition
is taken into account, give the same kind of results~: strong peaks appear and a
single one dominated the others as time increases (due to finite size effects).
It has to be notice that the flux of particles which fall down the edges of the
column formed by the discrete model is not equivalent to a relaxation for the
continuous one. This last term has a smooth effect. Nevertheless it suggests to
introduce an anisotropic diffusion.\\

To conclude, we have proposed a new continuous model for which
main ingredients are a non-linear shadowing deposit (proportional to the square of
the local exposure angle $\Omega$) and an anisotropic diffusion. The numerical
simulation results indeed show the formation of height columns, with sharp edges.
Furthermore, numerical simulations show that it is necessary to deal with a nonlinear shadowing
to obtain columnar shape deposit.\\

\end{document}